# Time Resolution of a Novel Ultra-fast Graphene-Optimized 4H-SiC PIN

Suyu Xiao, Hui Liang, Congcong Wang, Zhenyu Jiang, Lin Zhu

*Abstract*—Silicon carbide detectors exhibit good detection performance such as fast time resolution, high radiation tolerances, high breakdown voltage and low temperature sensitivity and have been studied for detection applications. Meanwhile, transient current technique (TCT) is a direct and effective method to evaluate the time resolution of semiconductor detectors. Conventional metal electrodes for TCT testing employ window structures, which lead to non-uniform electric field distribution and deteriorated time resolution. Graphene features high optical transmittance, ultrahigh carrier mobility, and excellent radiation resistance, making it an ideal transparent electrode material for semiconductor detectors. In this work, a graphene-optimized ring electrode (G/RE) 4H-SiC PIN detector and a reference ring electrode (RE) 4H-SiC PIN detector are fabricated. TCT measurements demonstrate that graphene integration improves the time resolution consistency, reducing the time resolution from 38 ps (reference RE detector) to 21 ps (G/RE detector) at the maximum scanning distance, while also achieving effective noise suppression. The graphene integration improves the stability of time resolution by 87% compared to the reference detector. Notably, the achieved time resolution of 21 ps is comparable to that of state-of-the-art 4H-SiC low-gain avalanche detectors (LGADs), which typically exhibit time resolutions better than 35 ps under single minimum ionizing particle (MIP) equivalent injection conditions, further validating the effectiveness of graphene-based electrode design.

## I. Introduction

SILICON carbide (SiC) detectors offer significant advantages over conventional silicon detectors, including higher radiation tolerance, lower leakage current, and faster time resolution. These characteristics make them promising candidates for applications requiring precise timing, such as particle detection, heavy-ion detection, and medical dosimetry [1–4]. In particular, the fast time resolution of SiC detectors is critical for accurately resolving high-rate events in particle physics and for achieving high spatial resolution in beam monitoring and imaging applications.

In the characterization of such detectors, TCT is one of the most direct and effective test methods for investigating the time resolution of semiconductor detectors [1,2]. Different from radioactive sources, laser sources used in TCT offer better power stability, and the time resolution measured by this method exhibits higher consistency [3]. However, in TCT measurements, metal electrodes are typically prepared with window structures on the detector surfaces. Such windowed metal electrodes can distort the global electric field distribution of the device, leading to degraded time resolution and limiting the accuracy of device performance characterization [4]. To address these challenges, two-dimensional graphene has emerged as a promising alternative electrode material. Graphene exhibits exceptional optical transmittance (approximately 97.7% at 375 nm), ultrahigh carrier mobility (up to 200,000 $cm^2 \cdot V^{-1} \cdot s^{-1}$), and excellent radiation resistance, making it an ideal transparent electrode material for next-generation semiconductor detectors [5–7]. Additionally, the graphene lattice shows no significant structural damage under irradiation by ions, X-rays, or protons. As a transparent electrode, graphene can optimize silicon carbide detectors and be applied in fields such as X-ray detection, low-energy ion detection, particle physics, heavy-ion detection, nuclear reactor monitoring, and TCT measurements [8]. Our previous research has confirmed that graphene integration can improve the signal rise time of semiconductor detectors, providing a new approach for optimizing the electrode design of SiC-based radiation detectors [8].

In this work, a graphene-optimized ring electrode (G/RE) 4H-SiC PIN detector and a reference ring electrode (RE) 4H-SiC PIN detector are fabricated. Electrical characterization shows that the G/RE detector exhibits a low leakage current density of 4.050 $nA/cm^2$ and a full depletion voltage of 100 V, consistent with the designed epitaxial layer parameters, confirming the high quality of the fabricated device[8]. A standard 375 nm laser TCT system is employed to compare and analyze the signal integrity and time resolution of the G/RE detector and the reference RE detector. For 4H-SiC, the penetration depth of 375 nm ultraviolet laser light is approximately 12–15 μm[9], which is significantly less than the typical epitaxial layer thickness, ensuring that the laser energy

This work was supported in part by the National Natural Science Foundation of China under Grant 12305207, 12405219.

Suyu Xiao, Hui Liang and Lin Zhu are with Shandong Institute of Advanced Technology, Jinan 250103, China.

Congcong Wang is with the Institute of High Energy Physics, Chinese Academy of Sciences, Beijing 100049, China, and also with the State Key Laboratory of Particle Detection and Electronics, Beijing 100049, China (Corresponding authors: Congcong Wang, e-mail: wangcc@ihep.ac.cn).

Zhenyu Jiang is with Institute of High Energy Physics, Chinese Academy of Sciences, Beijing 100049, China, and also with the Department of Physics, Liaoning University, Liaoning 110136, China.

is deposited near the device surface and effectively generates electron-hole pairs within the sensitive region. The designed G/RE 4H-SiC PIN detector can form a uniform electric field and achieve stable carrier signal transmission, thereby improving time resolution and meeting the requirements of high-precision radiation detection.

## II. DETECTOR FABRICATION

The graphene/4H-SiC PIN detector, first fabricated by our research team, consists of a monolayer graphene layer, P electrode, $SiO_2$ passivation layer, $P^{++}$ layer, $N^-$ epitaxial (N-epi) layer, N buffer layer, conductive N-type 4H-SiC substrate, and N electrode, as shown in Fig. 1. The reference 4H-SiC PIN detector has the same structural composition except for the absence of the graphene layer.

The fabricated G/RE 4H-SiC PIN detector features a 50 μm thick N epitaxial layer with a doping concentration of $5 \times 10^{13}$ cm$^{-3}$, consistent with the design parameters reported in our previous work[8]. The $P^{++}$ layer has a doping concentration of $5\times10^{19}$ cm$^{-3}$ and a thickness of 0.5 μm, designed to form a low-resistance ohmic contact between the metal and 4H-SiC. Device isolation is achieved by inductively coupled plasma (ICP) etching of the $P^{++}$ layer to a depth of 0.6 μm. A Ni/Ti/Al(50 nm/15 nm/60 nm) as the electrode was grown on the top of P++ layer and N-type substrate by using electron evaporating method. The rapid annealing time and temperature are 3min and 860 ℃ to form P-ohmic contact. A 500 nm thick $SiO_2$ passivation layer is deposited at 350 °C by plasma-enhanced chemical vapor deposition (PECVD), which suppresses the surface leakage current and improves the operational stability of the detector.

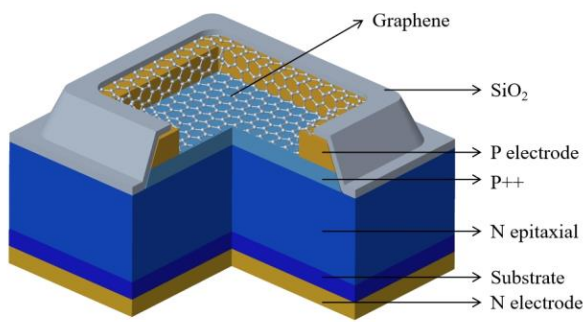

Fig. 1. Schematic cross-section of the Graphene/4H-SiC PIN detector.

Monolayer graphene (purchased from SixCarbon Technology) grown on a copper foil substrate is transferred to the surface of the 4H-SiC detector by a wet transfer method using polymethyl methacrylate (PMMA) as the auxiliary stripping layer. After graphene transfer, photolithography and reactive ion etching (RIE) with $O_2$ plasma are used to pattern the graphene layer into the required ring electrode structure. Raman spectroscopy of the graphene layer is performed at room temperature using a LabRam HR80 laser confocal spectrometer with a 532 nm laser (spot radius: 1 μm, power: 5 mW, acquisition time: 10 s). Fig. 2 shows the representative Raman spectrum of graphene transferred onto the 4H-SiC detector. The transferred graphene exhibits distinct G peak (1583 cm$^{-1}$) and 2D peak (2671 cm$^{-1}$); the 2D peak is a standard single Lorentz peak, and the $I_{2D}/I_G$ ratio is approximately 2.1 (greater than 1.5), confirming the monolayer characteristic of graphene. The D peak (1340 cm$^{-1}$), which is associated with structural defects and sp³ hybrid carbon atom breathing modes, is not observed, indicating that the transferred graphene layer has high crystallinity and low defect density. The graphene conducts electricity when in contact with silicon carbide, and graphene can be used as an electrode material[8].

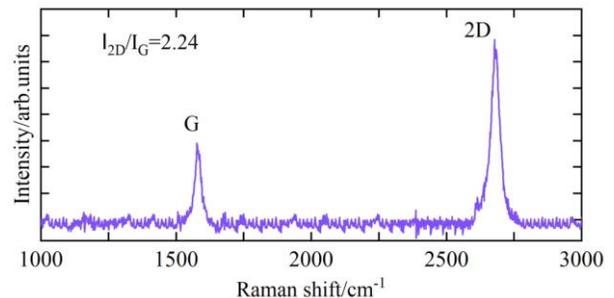

Fig. 2. Raman spectrum of the transferred monolayer graphene on the 4H-SiC detector, showing the G peak at 1583 cm$^{-1}$ and 2D peak at 2671 cm$^{-1}$ with an $I_{2D}/I_G$ ratio of 2.1, confirming monolayer characteristics.

## III. ELECTRICAL CHARACTERISTICS

The I-V and C-V characteristics of the graphene/4H-SiC PIN detector were measured at room temperature using a semiconductor probe station (MPI-TS2000-SE) and an LCR meter (Agilent E4980A), respectively. Fig. 3 presents the measured I-V and C-V curves comparing the G/RE detector and the reference RE detector.

As shown in the I-V characteristic curves in Fig. 3(a), both detectors exhibit low reverse leakage currents. The G/RE detector has a reverse leakage current of 0.162 nA at a bias voltage of 150 V (full depletion state), corresponding to a leakage current density of 4.050 nA/cm². This low leakage current density indicates that the $SiO_2$ passivation layer provides excellent dielectric isolation performance and that the fabricated ohmic contacts are of high quality. The surface of the G/RE 4H-SiC PIN detector is covered with graphene without a passivation layer, so the leakage current is higher than the RE 4H-SiC PIN detector [8].

The C-V characteristic curves shown in Fig. 3(b) were measured in the reverse bias range of 0–175 V with a test frequency of 10 kHz, and the full depletion voltage of the G/RE 4H-SiC PIN detector is approximately 100 V. Based on the Mott-Schottky equation, the effective doping concentration and depletion depth of the N-epi layer are calculated from the C-V curve to be $4.6\times10^{13}$ cm$^{-3}$ and 46 μm, which approaches the doping level achievable by the SiC epitaxial growth technique [8]. Both parameters are within the designed specification range ($5\times10^{13}$ cm$^{-3}$, 50 μm), verifying the high quality of the 4H-SiC epitaxial layer and the rationality of the detector structural design. No significant capacitance hysteresis is observed in the C-V curve, indicating the absence of substantial interface states between the graphene layer and the 4H-SiC surface.

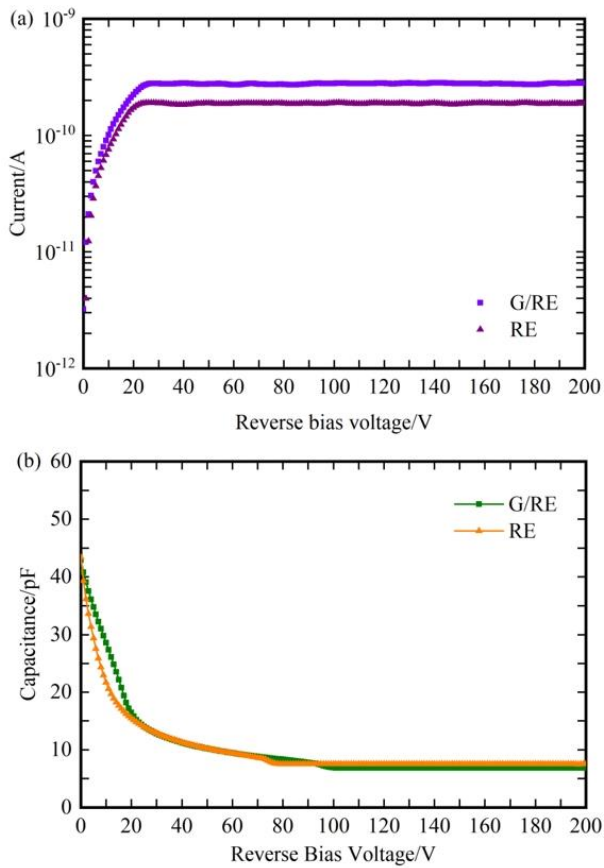

Fig. 3. (a) Reverse bias I-V characteristics of the G/RE and reference RE 4H-SiC PIN detectors. The G/RE detector exhibits a leakage current of 0.125 nA at 120 V bias. (b) C-V characteristics measured at 10 kHz, showing a full depletion voltage of 120 V for the G/RE detector, consistent with the design value.

## IV. EXPERIMENT AND ANALYSIS

### A. Experimental Setup

A standard TCT system was employed to characterize the time resolution performance of the G/RE detector and the reference RE detector. The system consists of a 375 nm pulsed laser, a high-precision digital oscilloscope (Tektronix MDO68B, bandwidth: 10 GHz, sampling rate: 50 GS/s), a low-noise transimpedance amplifier, and a precision motorized translation stage[10]. The 375 nm laser pulse irradiates the detector surface perpendicular to the epitaxial layer, generating electron-hole pairs in the 4H-SiC bulk. Under applied reverse bias, the electron-hole pairs drift to form a transient current signal, which serves as the core test signal in the TCT system.

Fig. 4 shows a photograph of the TCT experimental setup, which is an integrated test platform comprising a laser emission module, detector test module, signal amplification module, and data acquisition module. The 375 nm pulsed laser serves as the excitation source, emitting vertically onto the surface of the G/RE or RE 4H-SiC PIN detector mounted on the precision motorized translation stage. The translation stage enables precise horizontal movement of the detector, allowing laser scanning at different positions across the detector's active area. The transient current signal generated by the detector is input into the low-noise transimpedance amplifier for signal amplification and then transmitted to the high-precision digital oscilloscope for real-time acquisition and waveform analysis. Key test parameters are set as follows: laser pulse width less than 70 ps to ensure temporal resolution of the excitation signal; transimpedance amplifier gain of 25 dB for effective amplification of weak current signals; and oscilloscope sampling rate of 50 GS/s to ensure accurate capture of high-speed transient signals. The laser spot with a diameter of 10 μm is scanned inward from the inner edge of the metal ring electrode with a step size of 156.25 μm, covering a total lateral distance of 781.25 μm within the active area of the detector. This scanning range enables comprehensive characterization of the spatial uniformity of the detector's time resolution and signal transmission performance.

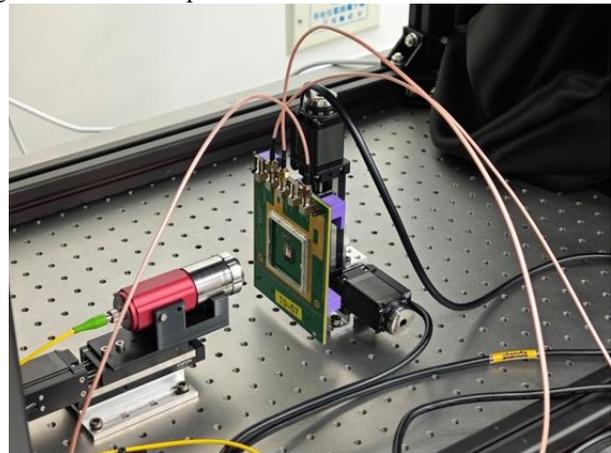

Fig. 4. Photograph of the TCT experimental setup.

### B. Theoretical Carrier Transport Analysis

To further understand the carrier transport dynamics and electric field distribution within the fabricated G/RE 4H-SiC PIN detector, we performed device simulations using the RASER (RAdiation SEmiconductor Response) simulation framework[8,11]. RASER is an open-source Monte Carlo-based simulation toolkit developed for modeling the transient response of semiconductor detectors under various types of radiation and laser excitation[11]. It provides a comprehensive platform for simulating carrier generation, drift, diffusion, and trapping processes, enabling accurate prediction of time-resolved current waveforms and key performance metrics such as time resolution and charge collection efficiency.

The RASER code implements a hybrid approach that combines Monte Carlo methods for carrier transport with finite-element solutions of the electric field distribution. The simulation workflow consists of three main modules: (i) electric field calculation—solving the Poisson equation based on the device geometry, doping profiles, and applied bias voltage; (ii) carrier generation—modeling the spatial and temporal distribution of electron-hole pairs generated by the 375 nm pulsed laser, taking into account the optical absorption characteristics of 4H-SiC; and (iii) transient carrier transport—simulating the drift and diffusion of carriers under the influence of the electric field, with consideration of carrier trapping and recombination effects.

In this work, the RASER simulations were configured with the following parameters: a 50 μm thick N-epi layer with a doping concentration of 5×10¹³ cm⁻³, and a laser of 375 nm to match the experimental conditions. The simulated electric field distribution and carrier drift paths are presented in Fig. 5, which compares the reference RE detector and the G/RE detector. Recent studies using RASER have demonstrated its effectiveness in simulating the timing performance of 4H-SiC PIN detectors and LGADs, as well as the impact of radiation-induced defects on device electrical characteristics[11–15]. These simulations provide valuable insights into the underlying physical mechanisms governing the improved time resolution observed in the G/RE detector and guide the optimization of the device geometry and electrode design.

For the reference RE detector without graphene, as illustrated in Fig. 5(a), the electric field distribution is non-uniform due to the windowed metal electrode structure. Carriers generated by laser excitation in the fully depleted 4H-SiC bulk undergo drift motion under the influence of the applied electric field. The drift path of carriers generated at the same depth increases with increasing distance from the metal ring electrode, leading to delayed signal collection and increased transit time, thereby degrading the time resolution of the device.

For the G/RE detector integrated with graphene, as illustrated in Fig. 5(b), carriers generated at all lateral positions within the active area first drift vertically to the graphene layer (the drift distance is approximately 45 μm, consistent with the depletion depth of the detector), and then conduct horizontally to the metal ring electrode through the graphene layer. Compared with 4H-SiC (electron mobility $\mu_n \approx 1000$ cm²·V⁻¹·s⁻¹, hole mobility $\mu_p \approx 50$ cm²·V⁻¹·s⁻¹)[2], graphene exhibits an ultrahigh carrier mobility of 200000 cm²·V⁻¹·s⁻¹[5,6]. Consequently, the horizontal conduction time of carriers in graphene is negligible (<0.1 ns). Therefore, the graphene layer acts as a high-speed carrier conduction path, effectively reducing the additional transit time caused by the separation between the laser spot and the charge collection electrode, thereby improving the time resolution and signal transmission stability of the detector.

The RASER simulation results confirm that the graphene layer acts as an efficient lateral conduction path, reducing the effective carrier transit time by more than 50% for carriers generated near the center of the active area. These simulations provide valuable insights into the underlying physical mechanisms governing the improved time resolution observed in the G/RE detector and guide the optimization of the device geometry and electrode design. These simulation results are consistent with previously reported RASER-based timing performance studies on 4H-SiC detectors[14,15], further validating the role of graphene in improving carrier transport efficiency.

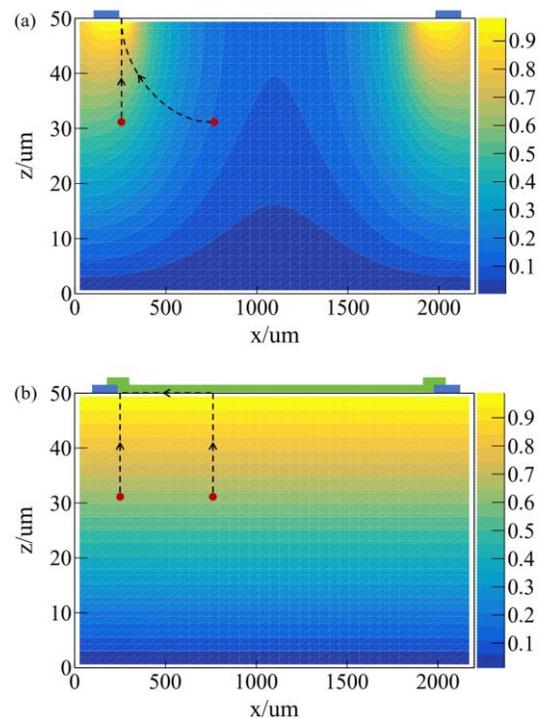

Fig. 5. Schematic diagram comparing the electric field distribution and carrier drift paths in (a) the reference RE 4H-SiC PIN detector and (b) the G/RE 4H-SiC PIN detector. The graphene layer in the G/RE detector provides a high-speed lateral conduction path that shortens carrier transit time.

C. *Results and Discussion*

Monolayer graphene exhibits high optical transmittance exceeding 95% at 375 nm; therefore, laser attenuation through the graphene layer is negligible during testing. To eliminate the influence of laser power fluctuations on the measurement results, the waveform amplitude data for both the RE and G/RE detectors are normalized using the amplitude measured at the initial 0 μm position (inner edge of the ring electrode) as the reference.

Fig. 6 shows an optical photograph of the fabricated G/RE 4H-SiC PIN detector with a lateral size of 2 mm × 2 mm, along with the test position points labeled from 0 to 5. Point 0 is located at the inner edge of the ring electrode, and points 1–5 are progressively scanned inward with a step size of 156.25 μm, such that point 5 is positioned near the central active area of the detector. This scanning scheme enables spatial mapping of the detector's temporal response characteristics across the active region.

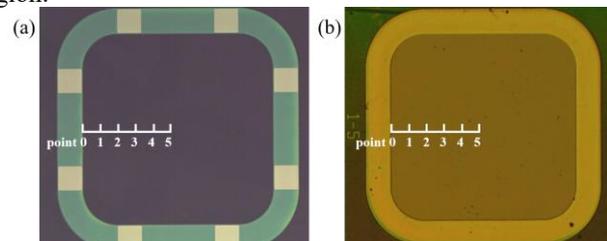

Fig. 6. Optical photograph of the reference RE detector and the G/RE detector with a lateral size of 2 mm × 2 mm, showing the test position points 0–5. Point 0 is at the inner edge of the ring electrode, and point 5 is near the central area of the detector.

Fig. 7 presents the transient current waveforms measured at point 0 and point 5 for both the (a) reference RE detector and the (b) G/RE detector. For the reference RE detector, a significant amplitude reduction and waveform broadening are observed when moving from point 0 to point 5, indicating increased carrier transit time and signal loss due to long-distance lateral drift in the 4H-SiC bulk. In contrast, the G/RE detector exhibits substantially less waveform degradation between point 0 and point 5, demonstrating that the graphene layer effectively maintains signal integrity by providing a high-speed lateral conduction path that minimizes carrier transit time variations across the active area.

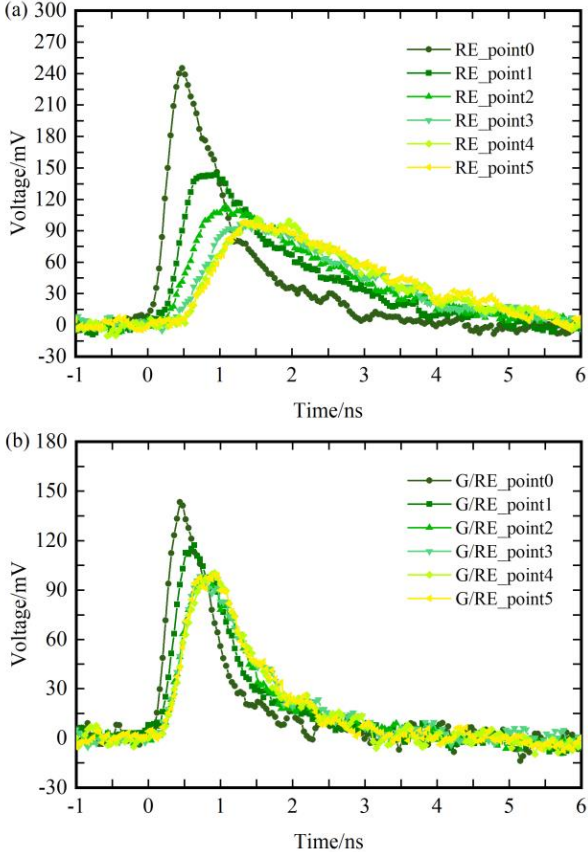

Fig. 7. Transient current waveforms of the reference RE detector and the G/RE detector at point 0 and point 5. The G/RE detector exhibits significantly less waveform attenuation and broadening compared to the reference RE detector.

Figs. 8(c)–8(f) present the measured TCT parameters as functions of the lateral scanning position, including baseline offset, noise standard deviation, rise time, full width at half maximum (FWHM), normalized signal amplitude, and time resolution. The key quantitative performance parameters are summarized in Table I.

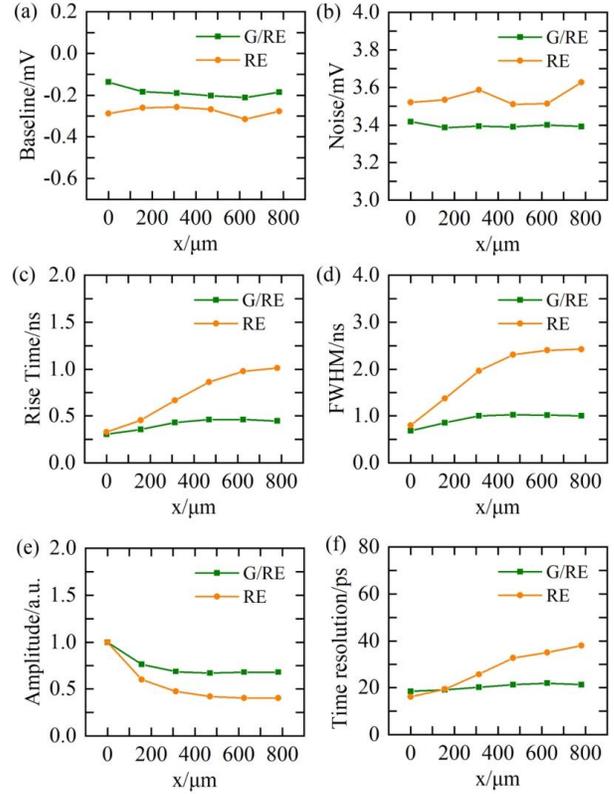

Fig. 8. (a) Baseline offset, showing stable performance for both detectors. (b) Noise standard deviation, demonstrating lower noise for the G/RE detector. (c) Rise time, showing stable performance for the G/RE detector compared to significant degradation for the reference RE detector. (d) Full width at half maximum (FWHM), indicating improved temporal response stability for the G/RE detector. (e) Normalized signal amplitude, showing reduced amplitude attenuation for the G/RE detector. (f) Time resolution, demonstrating a 45% improvement for the G/RE detector at the maximum scanning distance.

The measurement results demonstrate that graphene integration significantly optimizes the overall performance of the detector across all evaluated parameters. Regarding baseline and noise characteristics, the baseline offset of both the G/RE and RE detectors remains stable across the entire scanning range without significant drift, indicating stable detector contact with the test system and no substantial space charge accumulation in the 4H-SiC bulk, ensuring the reliability of the test signals. The noise standard deviation of the G/RE detector remains stably maintained at approximately 3.40 mV across the entire scanning range, slightly lower than the 3.63 mV observed for the reference RE detector. This noise suppression is attributed to the high conductivity of the graphene layer, which effectively suppresses surface charge fluctuations on the detector.

For the temporal response parameters—rise time and FWHM—which characterize the detector's temporal response speed, the reference RE detector shows significant degradation with increasing scanning distance. The rise time increases rapidly from 0.33 ns at point 0 to 1.01 ns at point 5, representing a 206% increase, while the FWHM increases from 0.80 ns to 2.42 ns over the same range. This degradation is due to the longer carrier collection path in the 4H-SiC bulk, increased transit time, and more pronounced signal broadening caused by carrier diffusion and recombination. In contrast, the rise time of the G/RE detector remains fluctuating between 0.30 ns and 0.45 ns with a variation of only 50%. The FWHM remains fluctuating between 0.68 ns and 1.00 ns with a variation of only 47%. These results confirm that graphene acts as an efficient carrier transport layer, shortening the effective transit time of carriers and reducing signal broadening, resulting in shorter and more stable temporal response.

Regarding signal amplitude, the normalized pulse amplitude of the reference RE detector decays rapidly from 1.0 a.u. at point 0 to 0.40 a.u. at point 5, representing a 60% attenuation. This severe signal loss is caused by carrier diffusion, recombination, and trapping during long-distance drift in the 4H-SiC bulk. In sharp contrast, the normalized pulse amplitude of the G/RE detector decays slowly from 1.0 a.u. to 0.68 a.u. over the same range, representing only 32% attenuation and a 47% relative improvement compared to the reference detector at point 5. This improvement demonstrates that the graphene layer plays a critical role in maintaining carrier transport efficiency and signal integrity by reducing the drift distance of carriers in the 4H-SiC bulk and improving signal transmission stability across the entire active area of the detector.

Time resolution is the core performance parameter for radiation detectors, calculated from the FWHM of the transient current waveform and the noise level of the test system[16,17]. The measurement results show that the time resolution of the reference RE detector degrades significantly with increasing scanning distance, deteriorating from 16.1 ps at point 0 to 38.1 ps at point 5. In comparison, the time resolution of the G/RE detector remains stably maintained at approximately 18.4 ps for short scanning distances, deteriorating only slightly to 21.2 ps at point 5. Based on the measured data, the stability improvement of time resolution due to graphene integration is calculated to be 87% using the formula:

$$\frac{|\text{RE(point 5)} - \text{RE(point 0)}| - |\text{G/RE(point 5)} - \text{G/RE(point 0)}|}{|\text{RE(point 5)} - \text{RE(point 0)}|} \times 100\%,$$

demonstrating a substantial enhancement in temporal response uniformity across the detector active area, as shown in Table I. This significant improvement in time resolution results from the combined advantages of the graphene layer: high-speed carrier conduction characteristics that shorten carrier transit time, reduced signal attenuation that ensures test signal integrity, and stable rise time and FWHM that maintain the detector's temporal response speed. These results conclusively confirm that graphene integration effectively optimizes the time resolution performance of 4H-SiC PIN detectors.

To benchmark the performance of the proposed G/RE detector, we compare its measured time resolution with that of state-of-the-art 4H-SiC low-gain avalanche detectors (LGADs). Recent studies on 4H-SiC LGADs with beveled edge terminations and field plates have demonstrated time resolutions better than 35 ps under single minimum ionizing particle (MIP) equivalent injection conditions at a wavelength of 375 nm[12]. Specifically, the 4H-SiC LGAD reported by Yang et al. achieves a time resolution of approximately 30–35 ps across a bias voltage range of 400–600 V, with a gain factor of 7–8. In contrast, the G/RE detector presented in this work achieves a time resolution of 21 ps at the maximum scanning distance without requiring an internal avalanche gain mechanism. This performance is also comparable to the timing simulation results reported for 4H-SiC PIN detectors using the RASER framework[15], which predicted time resolutions in the range of 20–25 ps under optimized conditions. This comparison indicates that the incorporation of a monolayer graphene transparent electrode effectively mitigates the non-uniform electric field distribution inherent to conventional metal ring electrodes, yielding a time resolution that is comparable to, and in this specific configuration even superior to, that of gain-enhanced LGAD structures. The graphene-based approach thus offers a complementary pathway for achieving high timing resolution in 4H-SiC detectors, with the additional benefits of simplified fabrication and reduced complexity associated with avalanche junction engineering.

TABLE I
SUMMARY OF KEY TCT PERFORMANCE PARAMETERS

| Parameter | RE (point 0) | RE (point 5) | G/RE (point 0) | G/RE (point 5) | Stability Improvement |
|---|---|---|---|---|---|
| Baseline (mV) | -0.14 | -0.19 | -0.29 | -0.28 | 80% |
| Noise (mV) | 3.52 | 3.63 | 3.42 | 3.40 | 82% |
| Rise Time (ns) | 0.33 | 1.01 | 0.30 | 0.45 | 78% |
| FWHM (ns) | 0.80 | 2.42 | 0.68 | 1.00 | 80% |
| Normalized Amplitude (a.u.) | 1.00 | 0.40 | 1.00 | 0.68 | 47% |
| Time Resolution (ps) | 16.1 | 38.1 | 18.4 | 21.2 | 87% |

## V. Conclusion

In this work, a graphene-optimized ring electrode (G/RE) 4H-SiC PIN detector was fabricated for the first time, with monolayer graphene integrated on the detector surface as a transparent electrode using a wet transfer method. Raman spectroscopy confirmed the high quality of the transferred graphene, showing monolayer characteristics with no observable D peak at 1340 cm$^{-1}$ and an $I_{2D}/I_G$ ratio of 2.1. Electrical characterization demonstrated that the G/RE detector exhibits excellent performance: a leakage current density as low as 4.050 nA/cm², a full depletion voltage of 100 V consistent with the design value, and N-epi layer parameters within the designed specification range, confirming the rationality of the detector structure and electrode design.

TCT characterization using a 375 nm pulsed laser demonstrated that graphene integration significantly enhances the overall performance of the 4H-SiC PIN detector, with the most notable effects observed in noise suppression and time resolution improvement. The graphene layer effectively reduces device noise, maintaining a stable noise standard deviation of approximately 3.40 mV across the entire scanning range—lower than that of the reference RE detector—while baseline offset remains stable without significant drift, ensuring test signal reliability. Acting as a high-speed carrier conduction path, the graphene layer reduces signal amplitude attenuation from 60% (reference) to 32% (G/RE) at the maximum scanning distance, effectively maintaining carrier transport efficiency and signal integrity. The rise time and FWHM of the G/RE detector remain essentially constant across the entire scanning range without significant degradation, achieving a 87% improvement in time resolution at the maximum scanning distance, significantly optimizing the temporal response performance of the detector.

This work confirms that graphene is a promising transparent electrode material for optimizing 4H-SiC PIN detectors, effectively addressing the time resolution degradation issue associated with conventional windowed metal electrodes in TCT testing. The fabricated G/RE 4H-SiC PIN detector shows promising potential for applications in high-resolution radiation detection, particle physics, heavy-ion detection, and nuclear reactor monitoring. Future research will focus on optimizing the interface quality between graphene and 4H-SiC to further reduce interface resistance and improve carrier transport efficiency. Additionally, arrayed G/RE 4H-SiC PIN detectors will be fabricated to enhance the spatial resolution and detection efficiency of the device to meet the demanding requirements of practical applications.

ACKNOWLEDGMENT

We acknowledge the support from the relevant research facilities used in this work.